\documentclass[twocolumn,prl, showpacs]{revtex4-1}
\usepackage{graphicx}
\usepackage{dcolumn}
\usepackage{bm}
\usepackage{color}

\begin{document}


\title{Finite-size effect of antiferromagnetic transition and electronic structure in LiFePO$_4$}

\author{G. J. Shu$^1$}
\author{M. W. Wu$^2$}
\author{F. C. Chou$^{1,3,4}$}
\email{fcchou@ntu.edu.tw}
\affiliation{
$^1$Center for Condensed Matter Sciences, National Taiwan University, Taipei 10617, Taiwan}
\affiliation{
$^2$Department of Materials Science and Engineering, National Formosa University, Huwei 63249, Taiwan}
\affiliation{
$^3$National Synchrotron Radiation Research Center, Hsinchu 30076, Taiwan}
\affiliation{
$^4$Center for Emerging Material and Advanced Devices, National Taiwan University, Taipei 10617, Taiwan}

\date{\today}

\begin{abstract}
The finite-size effect on the antiferromagnetic (AF) transition and electronic configuration of iron has been observed in LiFePO$_4$.  Determination  of the scaling behavior of the AF transition temperature ($T_N$) vs. the particle size dimension ($L$) in the critical regime, $1-\frac{T_N(L)}{T_N(\infty)}\sim L^{-1}$, reveals that the activation nature of the AF ordering strongly depends on the surface energy.  In addition, the effective magnetic moment that reflects the electronic configuration of iron in LiFePO$_4$ is found sensitive to the particle size.  An alternative structural view based on the polyatomic ion groups of PO$_4^{3-}$ is proposed.  
\end{abstract}


                             
\maketitle


\begin{figure}
\includegraphics [width=3.5in] {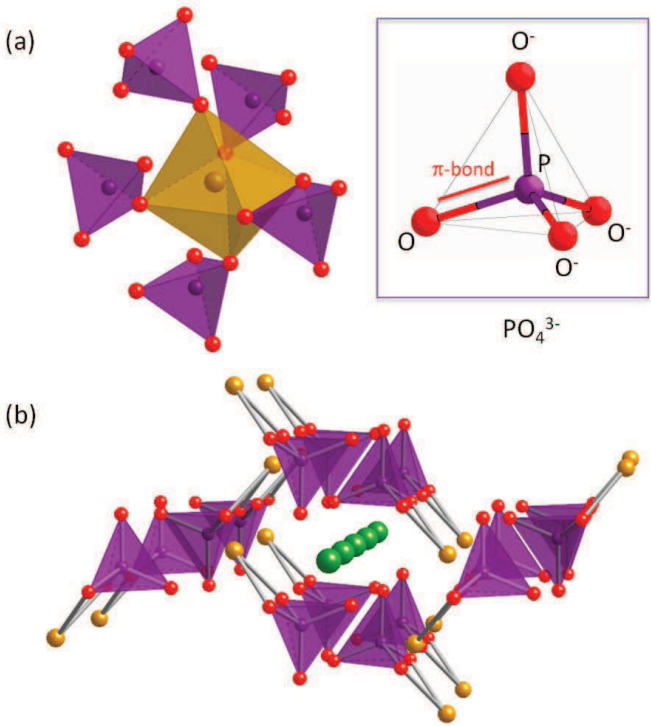}
\caption{\label{fig:fig-structure}(color online) (a) Iron ion environment of LiFePO$_4$ with PO$_4^{3-}$ ligand formed weaker crystal field.  The PO$_4^{3-}$ polyatomic ion is constructed by four $\sigma$-bond and one resonating $\pi$-bond as shown on the right.  (b) Alternative view of LiFePO$_4$ with lithium residing in the diffusion channels along \textbf{b}-direction of large pore size, which has van der Waals-like bonding with the neighboring Fe-PO$_4$ groups.}
\vspace{-5mm}
\end{figure}


LiFePO$_4$ has been considered as an ideal cathode electrode material with a workable redox potential of 3.45 V and a high theoretical capacity of 170 mAh/g.\cite{Padhi1997}  LiFePO$_4$ is a band insulator in bulk crystal form in agreement with the calculated large band gap,\cite{Zhou2004} however, the fast electron/ion charge/discharge process has been shown to be possible only when the particle size is reduced to below $\sim$100 nm.\cite{Yamada2001}  While significant progress has been made to demonstrate the impact of nano-scaling on defect chemistry, electrochemical behavior, electron-ion conductivity, lithium ion diffusion, and the charge/discharge rate in general,\cite{Meethong2007, Delmas2008, Julien2011} the fundamental question of ``why nano-scaling matters?" remains.     
Generally, there have been two major approaches to interpret the property improvement for LiFePO$_4$ using nano-scaling induced electron/ion transport: the first is called the "domino-cascade" model, which describes a fast anisotropic lithium insertion/extraction action coupled to the LiFePO$_4$-FePO$_4$ interface movement through an electron polaronic hopping mechanism in nano-crystallites,\cite{Delmas2008} and the second focuses on the surface effect with carbon coating and miscibility gap size reduction as a result of nano-scaling.\cite{ Meethong2007, Zaghib2010}


Conventionally, the iron ion has been viewed as Fe$^{2+}$ with oxygen ligands to form the octahedral crystal electric field (CEF) of e$_g$-t$_{2g}$ splitting.\cite{Zaghib2007}  However, the magnitude of the effective magnetic moment ($\mu_{eff}$) obtained from the Curie-Weiss law fitting was inconsistent and spread in the range of $\mu_{eff}$$\sim$4.7-5.5 $\mu_B$, which has been interpreted that Fe$^{2+}$ must be in the high-spin state (HS) as t$_{2g}^4$e$_g^2$ of S=2 with a spin-only value of $\mu_{eff}$ = 4.9$\mu_B$, and the inconsistent moment above 4.9 $\mu_B$ has often been described by incomplete orbital-quenching, i.e., the theoretically calculated value from the Lande g-factor with J=L+S (S=2, L=2) without orbital quenching should be $\mu_{eff}$=6.7$\mu_B$.\cite{Ashcroft, Chernova2011}  Considering a typical iron ion in the crystal field composed of oxygen ligands with high electronegativity, the HS choice of d$^6$ implies an unexpectedly small crystal field splitting between $e_g-t_{2g}$, unless the influence of the nearby PO$_4^{3-}$ phosphate groups are included.  In addition, within the explanation of incomplete orbital-quenching, the actual mechanism that triggers the different degree of orbital-quenching remains unknown.  Several alternative but inconclusive explanations for the excess moment have been suggested, including the existence of the FePO$_4$ phase, hidden impurity phases with Fe$^{3+}$, small magnetic polarons of large polarized macro-spin,\cite{Mauger2008} and the surface disorder-generated low-spin state (S=1/2) of Fe$^{3+}$.\cite{Zaghib2008}  

In the discussion of the spin state of the iron ion, we find the traditional view starting from the distorted FeO$_6$ octahedral crystal electric field cannot avoid the influence of the nearby phosphate group of PO$_4^{3-}$.  Alternatively, weaker crystal field surround Fe can be formed by the weak ligands of   (PO$_4$)$^{3-}$ as polyatomic ions as shown in Fig.~\ref{fig:fig-structure}(a).  Owing to the extremely high positive charge density of hypervalent P$^{5+}$ with a particularly small ion radius (0.17$\AA$ (IV)), the (PO$_4$)$^{3-}$ group has often been treated as one polyatomic ion of tetrahedral coordination in chemical reactions.  The standard view of the phosphate group PO$_4^{3-}$ is shown in Fig.~\ref{fig:fig-structure}(a), which is composed of P-O tetrahedral-shaped $sp^3d$ hybrid orbitals with four $\sigma$-bonds plus one resonating $\pi$-bond.  

Considering (PO$_4$)$^{3-}$ as a unit, the larger effective radius leads to a much lower average negative charge density per phosphate than pure O$^{2-}$.  The actual environment surrounding the iron ion can be described as either an edge-sharing Fe-(PO$_4$) pair, or an Fe-(PO$_4$)$_5$ crystal field as shown in Fig.~\ref{fig:fig-structure}.  While the neighboring oxygen atoms of high negative charge density for FeO$_6$ octahedron lead to strong CEF splitting of e$_g$-t$_{2g}$ $\sim$ 1-2 eV, the CEF constructed by the five phosphate groups of much lower electronic charge density and longer average Fe-(PO$_4$) distance would create a weak crystal field, which concurs with the experimental finding of the HS state choice for Fe$^{2+}$ of d$^6$.  The 1D diffusion channels along the \textbf{b}-direction of large pore size for lithium could prefer van der Waals-like bonding as depicted in Fig.\ref{fig:fig-structure}(b), similar to that of the typical battery cathode Li$_x$CoO$_2$ with lithium in the van der Waals gap between 2D CoO$_2$ layers.  


A change at the electronic configuration level allows LiFePO$_4$ to transform from a bulk band insulator to a good conductor through nano-scaling alone. To explore this mechanism using the newly proposed structural view based on the polyatomic ion unit of PO$_4^{3-}$, we report a complete magnetic property study of LiFePO$_4$ with controlled particle size, starting from a bulk single crystal to nano-scaled particles, of $\sim$50 nm level, in discrete steps.  The effective magnetic moment of iron has been found to depend on the particle size.  The effective magnetic moment changes as a result of the increase in surface energy, which implies that a subtle electronic structure change for iron occurs. This result could be closely related to the observed significant improvement of electronic conduction for the cathode
material LiFePO$_4$, which is composed of nano-scaled grains.\cite{Delmas2008}  In addition, we find that the antiferromagnetic (AF) phase transition temperature $T_N$ in the critical regime shows a scaling behavior in agreement with that predicted by the finite-size effect.  The particle size dependence of the electronic activation type has been established rigorously for both the paramagnetic spin state and the critical phenomenon for AF ordering. 

\begin{figure}
\begin{center}
\includegraphics [width=3.5in] {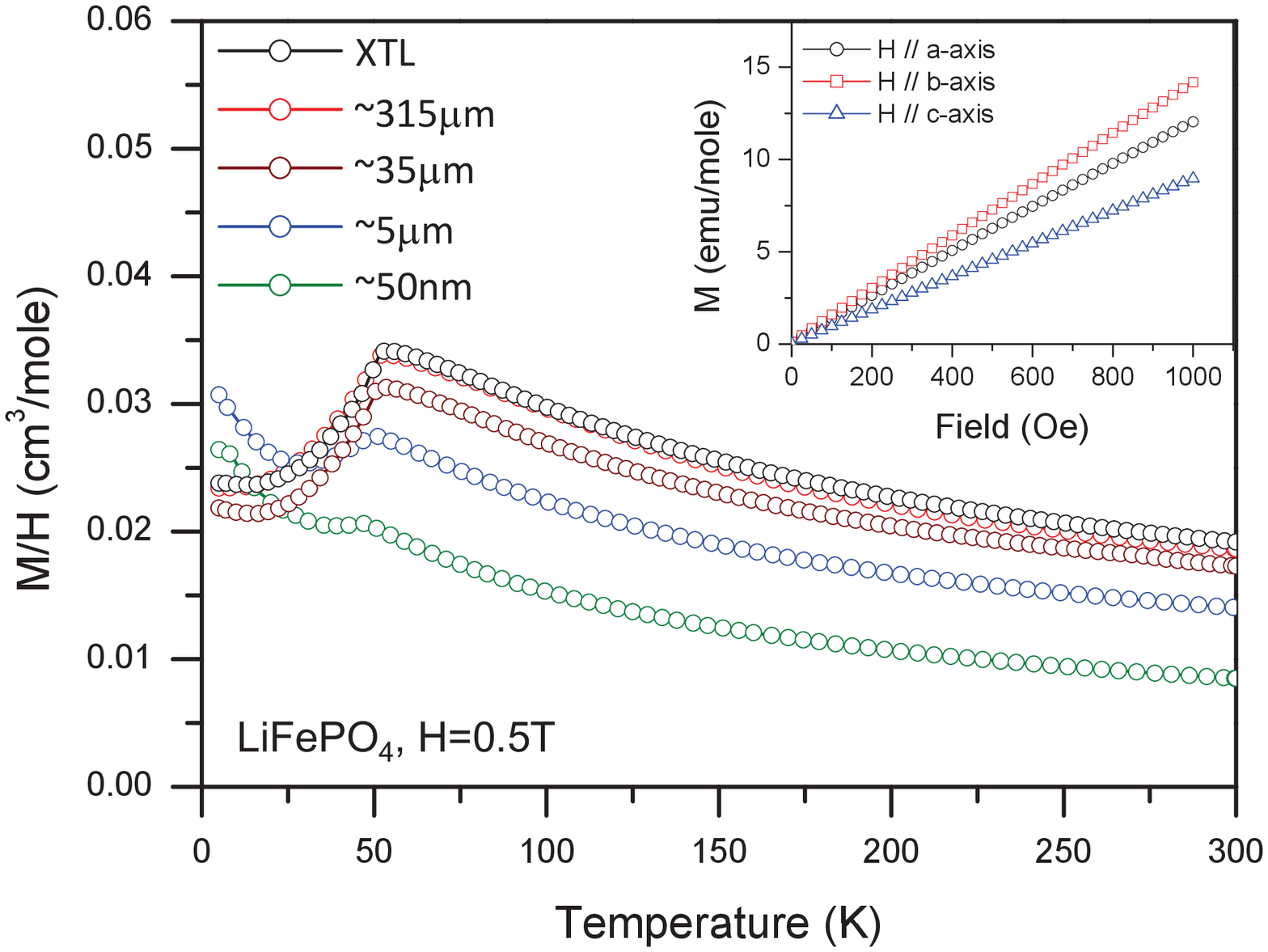}
\end{center}
\caption{\label{fig:fig-chiTH}(color online)  (a) Average spin susceptibilities as a function of temperature for various particle sizes. The single crystal data were obtained from the powder average of the oriented measurements as described in the text.  The isothermal magnetizations M(H) at 300 K for the single crystal along three major orientations are shown in the inset, no spontaneous  moment from FM impurity is found near the zero field. }
\end{figure}

A series of samples with different particle sizes (from bulk single crystal to $\sim$50 nm nano-crystallites) was prepared from one crystal batch using the jet-milling method. The experimental details of crystal growth, particle size reduction method, effective magnetic moment analysis, and particle size distribution are described elsewhere.\cite{Shu2012}  Chemical analysis using the ICP method has confirmed the Li:Fe ratio to be 1.002(5):1.  The average spin susceptibilities (M/H) as a function of temperature for various particle sizes are shown in Fig.~\ref{fig:fig-chiTH}(a).  The effective magnetic moments ($\mu_{eff}$) were extracted from the Curie-Weiss law fitting in the paramagnetic regime (100-300K) as $\chi$ = $\chi_\circ$+C/(T-$\Theta$) directly with the Curie constant C = N$_\circ$$\mu_{eff}^2$/3k$_B$.  Although Curie constant can also be extracted from the conventional approximation using 1/$\chi(T)$ $vs.$ T linear fitting, the nonlinearity introduced by the no-longer-negligible $\chi_\circ$ contribution at high temperature can be avoided.  In particular, to maintain a strict particle size dependence analysis of $\mu_{eff}$, the bulk single crystal sample has not been crushed into a powder, and its Curie constant has been extracted directly from the anisotropic spin susceptibilities along three major orthogonal axes following the equations derived by Liang \textit{et al.}\cite{Liang2008}

\begin{figure}
\includegraphics [width=3.5in]{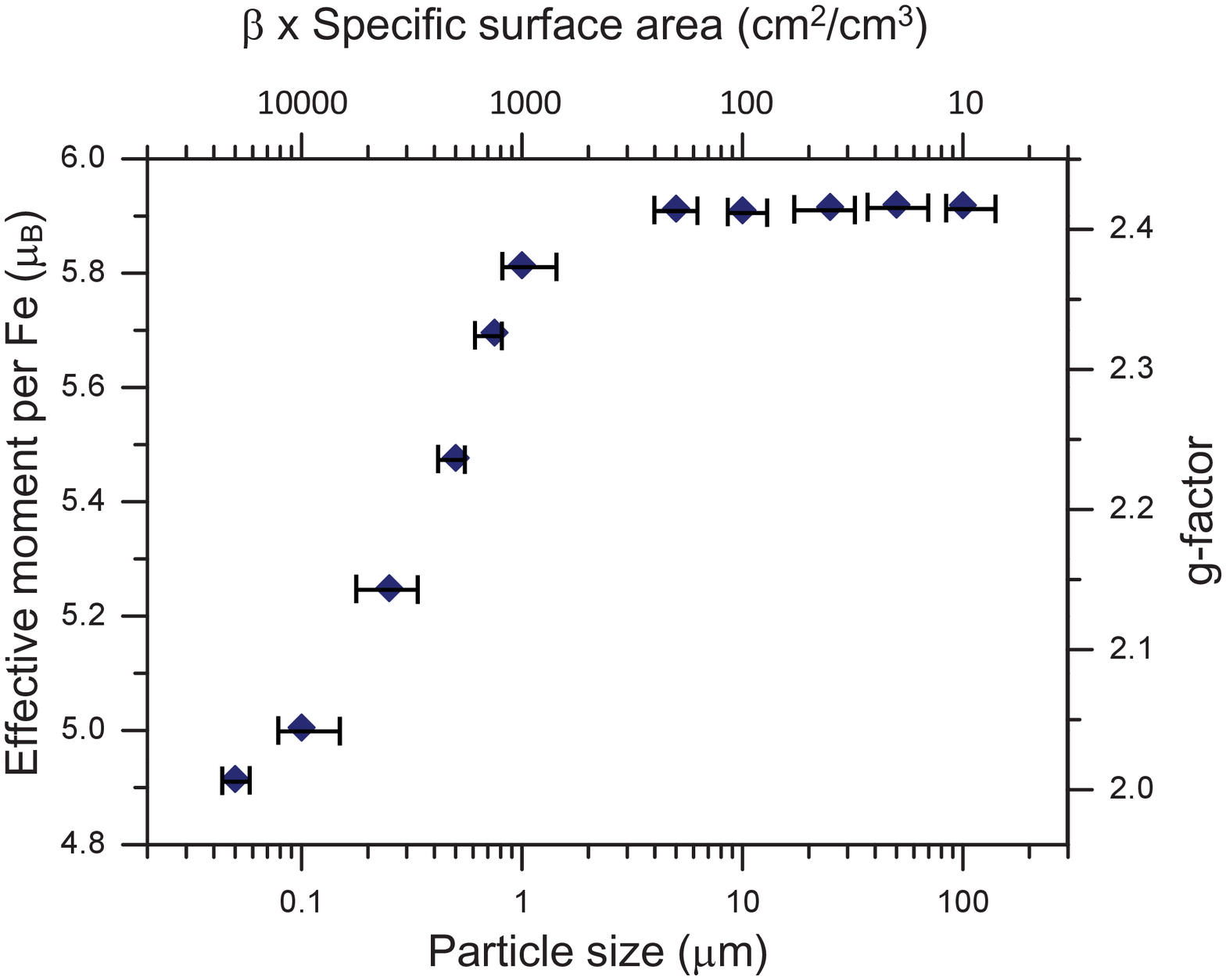}
\caption{\label{fig:fig-mueff-g}(color online) Particle size dependence of the effective magnetic moment $\mu_{eff}$, where the $\mu_{eff}$ values are derived from the fitted Curie constants using the Curie-Weiss law of M/H(T) shown in Fig.~\ref{fig:fig-chiTH}.  The upper x-axis shows the corresponding specific surface area with an undetermined geometric factor of $\beta$ ($\beta$=6 for cubic).  The corresponding g-factors are estimated by assuming $\mu_{eff}$=4.9$\mu_B$ of S=2. }
\vspace{-5mm}
\end{figure}

The effective magnetic moment $vs.$ particle size is summarized in Fig.~\ref{fig:fig-mueff-g}, which has been analyzed from the average spin susceptibility M/H(T) (Fig.~\ref{fig:fig-chiTH}) with an applied field of 0.5 Tesla.  It is clearly demonstrated that $\mu_{eff}$ for the bulk single crystal is $\sim$5.9 $\mu_B$ and this value monotonically reduces to $\sim$4.9 $\mu_B$ for the nano-size ($\sim$50 nm) particles.  
The HS-like configuration for Fe$^{2+}$ of d$^6$ is configured as $e_g^4t_{2g}^2$ with S=2 of $\mu_{eff}$=4.9 $\mu_B$ (assuming g=2).  Because $\mu_{eff}$=4.9 $\mu_B$ and $\mu_{eff}$=6.7 $\mu_B$ represent the theoretically calculated values using J=S and J=L+S (with L=S=2), respectively,\cite{Ashcroft} the $\mu_{eff}$ values larger than 4.90 $\mu_B$ could tentatively be interpreted as a result of the incomplete orbital-quenching.  Alternatively, we may infer the change in $\mu_{eff}$ from changing the g-factor beyond the electron g-factor of 2 as shown in Fig.~\ref{fig:fig-mueff-g}, which could be described as a result of change in local field.  A monotonic particle size dependence of $\mu_{eff}$ has been constructed as shown above, which implies that the earlier reports of inconsistent $\mu_{eff}$ values are closely related to the particle size.  However, the following questions arise: why does the perfect spin-only value of $\mu_{eff}$=4.9 $\mu_B$ exist only in the nano-scaled particles, and why is $\mu_{eff}$ higher in the single crystal sample?  Above all, whether this phenomenon is described by incomplete orbital-quenching or local field change, there must be an intrinsic electronic structure change as a function of particle size.


As shown in the inset of Fig.~\ref{fig:fig-chiTH}, the isotherm M(H) data at 300K have completely ruled out the existence of ferromagnetic impurity due to the absence of spontaneous magnetization jump near the zero field.  We find that the temperature-independent part of the average spin susceptibilities ($\chi_\circ$) is still at least one order higher than that estimated from the core diamagnetic, the Van Vleck paramagnetic, and the Pauli paramagnetic contributions combined.\cite{Johnston2010}   In addition, $\chi_\circ$ level decreases monotonically with the particle size as indicated in Fig.~\ref{fig:fig-chiTH}, which clearly opposes the assumption of higher  itinerant carrier densities for smaller particle size, i.e., a higher $\chi^{Pauli}$ contribution due to the increased carrier density of states near the Fermi level.  These results strongly suggest that a possible spin state change may have happened following the particle size change.  

\begin{figure}
\includegraphics [width=3.5in]{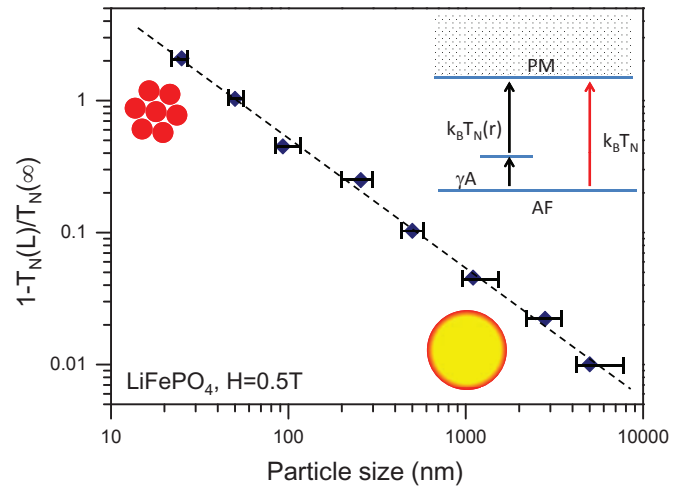}
\caption{\label{fig:fig-TN}(color online) The scaling behavior of reduced temperature $vs.$ dimension is plotted in the log-log scale, and the critical exponent $\nu$ for $1-\frac{T_N(L)}{T_N(\infty)}\sim L^{-\frac{1}{\nu}}$ is found to be close to 1.  For the large and nano-size particles, the high surface energy state is shown in red, and the ground state is shown in yellow.   The inset shows the AF spin gap and its relationship to the surface energy, as described in the text.}
\vspace{-5mm}
\end{figure}

In addition to the particle size dependence of the effective moment in the paramagnetic state, T$_N$ has shown an insignificant but consistent reduction for smaller particle sizes as revealed in Fig.~\ref{fig:fig-chiTH}(a), with T$_N$ $\sim$52 K for the single crystal and $\sim$49 K for the nano particle of $\sim$50 nm.  We find that the relationship between T$_N$ and the particle size can be scaled nicely by the reduced temperature $1-\frac{T_N(L)}{T_N(\infty)}$ $vs.$ particle size dimension L, where $T_N(\infty)$ corresponds to the single crystal sample of assumed infinite dimension, as  
\begin{equation}
1-\frac{T_N(L)}{T_N(\infty)}\sim L^{-\frac{1}{\nu}},
\end{equation}
\noindent and $\nu$ is fitted to be near 1 as plotted in Fig.~\ref{fig:fig-TN}.  Such scaling behavior perfectly demonstrates the finite-size effect when the magnetic correlation length is limited by the system dimension due to the effect of geometric confinement.\cite{Fisher1972, Batlle2002}  The current extracted critical exponent $\nu$$\sim$1 agrees well with the mean field theory prediction of a 3D Heisenberg system at the critical regime.

We find that the observed scaling behavior of the critical exponent $\nu$$\sim$1 affects the role of surface energy in a nano system.  Let us consider a single crystal sample of volume V, which is reduced into nano size particles of spherical shape with radius $r$ directly through jet milling particle size reduction.  We also assume that the observed reduction of T$_N(r)$ at nano size is due to the logarithmically enhanced surface energy of $\gamma A$ (A=total surface area) from the bulk ($\gamma$A$\sim$0) of a well-defined T$_N$, i.e., it requires less thermal energy $k_BT_N(r)$ to cross the AF spin gap ($k_BT_N$) in the critical regime, as illustrated in the inset of Fig.~\ref{fig:fig-TN}.   Then we can derive the scaling behavior directly in the mean field approximation as
\begin{equation}
k_BT_N - k_BT_N(r) \sim\gamma A = \gamma[(\frac{V}{\frac{4}{3}\pi r^3})(4\pi r^2)]\sim r^{-1},
\end{equation}   
\noindent which is comparable to the dimensionless scaling behavior of the critical exponent $\nu$=1 as shown above.  In the critical regime of AF transition, the correlation length of the order parameter, i.e., the stagger magnetization of the AF ordering, would diverge logarithmically with the reduced temperature.  Previously, the critical phenomenon was revealed by the scaling behavior using the magnetic correlation length ($\xi$).  Now, this phenomenon can be  explained intuitively using the role of surface energy in the nano system, i.e., the increased surface energy plays an equally important role as the thermal energy in a nano system, which should in turn adopt a different electronic structure at a raised ground state higher than that of a bulk single crystal, as reflected clearly by the changing $\mu_{eff}$ values as shown in Fig.~\ref{fig:fig-mueff-g}.

In summary, by carefully analyzing the effective magnetic moment of samples from single crystal to nano-size of $\sim$50 nm, we find that the finite-size effect on the transition temperature and the effective moment of LiFePO$_4$ reflect the significantly enhanced surface energy.
The observed size dependence of $\mu_{eff}$ should be viewed as a reflection of the change in the electronic structure as a result of competing thermodynamic factors to lower the total Gibbs free energy, i.e., the surface energy $\gamma \Delta A$, the entropy $T\Delta S$, and the magnetic energy $H \Delta M$ in $\Delta G = \Delta \textbf{H} - T\Delta S + \gamma \Delta A - H \Delta M$.   Each term in the total Gibbs free energy should be given nearly equal weight at the nano scale to minimize $\Delta$G in an intricate and competitive manner, instead of the enthalpy change dominating the minimization, which only reflects the ideal electronic structure.  


\section*{Acknowledgments}
FCC acknowledges the support from NSC-Taiwan under project number NSC 100-2119-M-002-021.  GJS acknowledges the support from NSC-Taiwan under project number NSC 100-2112-M-002-001-MY3.

\end{document}